\documentclass[11pt,a4paper]{article}
\usepackage[T1]{fontenc}
\usepackage{lmodern,amsmath,amssymb,graphicx,booktabs,tabularx,array,xurl,flafter}
\usepackage[top=24mm,bottom=24mm,left=28mm,right=28mm]{geometry}
\usepackage[authoryear,round]{natbib}
\setcitestyle{aysep={}}
\usepackage{microtype,caption,xcolor}
\definecolor{inklink}{RGB}{35,65,91}
\usepackage[colorlinks=true,allcolors=inklink]{hyperref}
\newcolumntype{Y}{>{\raggedright\arraybackslash}X}
\input{glyphtounicode}
\hypersetup{pdftitle={What Do Explanations and Fixes Add to LLM Code Review? A Replication on Audited Benchmarks},pdfauthor={Sabry E. Farrag}}
\usepackage{titlesec,fancyhdr,longtable,fancyvrb,textcomp,float,pdflscape}
\definecolor{accent}{HTML}{1B4F72}
\definecolor{muted}{HTML}{667889}
\titleformat{\section}{\sffamily\Large\bfseries\color{accent}}{\thesection}{.65em}{}
\titleformat{\subsection}{\sffamily\large\bfseries\color{accent}}{\thesubsection}{.65em}{}
\titlespacing*{\section}{0pt}{15pt}{6pt}
\titlespacing*{\subsection}{0pt}{11pt}{4pt}
\newcommand{\orcidmark}[1]{\href{https://orcid.org/#1}{\raisebox{-.17ex}{\includegraphics[height=1.62ex]{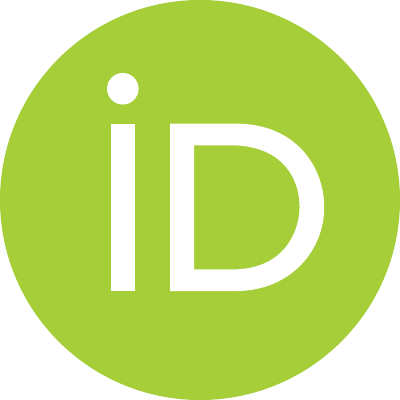}}}}
\begin{document}
\thispagestyle{plain}
\begin{center}
{\sffamily\fontsize{22}{25}\selectfont\bfseries\color{accent}What Do Explanations and Fixes Add to LLM Code Review? A Replication on Audited Benchmarks\par}
\vspace{10pt}
{\large Sabry E. Farrag\,\orcidmark{0009-0002-0735-7443}\par}
\vspace{5pt}
{\small School of Architecture, Computing and Engineering\\
University of East London, London, United Kingdom\par}
\vspace{3pt}
{\small Corresponding author: \href{mailto:sabryelgatem@gmail.com}{sabryelgatem@gmail.com}\par}
\end{center}
\vspace{3pt}
{\color{muted}\hrule}\vspace{8pt}
\noindent{\sffamily\bfseries\color{accent}Abstract}\par\nobreak\smallskip
\noindent Jin and Chen reported in Automated Software Engineering that asking language models to explain and fix code led them to reject substantially more correct programs. We replicated the study with six models reviewing short Python and Java programs. On the audited benchmarks, asking for an explanation and a fix raised false rejection by 5.4 percentage points over a verdict alone; its 95\% interval includes the five-point increase we had specified as practically important. The increase was smaller than in the original study. Using the original scoring, our reviewers rejected 20.3\% of programs labelled correct under the explain-and-fix prompt, compared with our average of 60.7\% from the original results. The audit found that the released verification code never called the assertions in the Mostly Basic Programming Problems benchmark. Subsequent counterexamples exposed defects in 79 programs that had passed the label audit. In a sensitivity analysis using adjudicated labels, false rejection was 3.7\% for verdicts alone and 5.9\% for explain-and-fix review. When reviewers correctly rejected buggy code under the richer prompt, their fixes passed the task's tests in 93.1\% of reviews. In an exploratory assessment, model judges found that 99.5\% of the explanations identified the planted defect. When reviewers rejected correct code, some proposed edits introduced failures. At the adjudicated rates, per 100 submissions with 10 buggy programs, explain-and-fix review would add about 2.0 false rejections and yield 9.2 fixes that pass the tests.\par
\medskip\noindent{\small\textbf{Keywords}\quad large language models, code review, LLM-as-a-judge, replication, benchmark label errors, requirement conformance\par}

\section{Introduction}

Claude Haiku rejected a QuixBugs function for computing the greatest common divisor. Its explanation traced \texttt{gcd(35, 21)} to \texttt{gcd(14, 21)} and showed that the next recursive call repeated the same arguments. Its fix moved the remainder to the second argument. The edited function returned 7 for the example and passed the QuixBugs tests. The developer could inspect the failing input, the faulty line and the proposed repair.

\begin{figure}[!tbp]
\centering
\includegraphics[width=\linewidth,keepaspectratio]{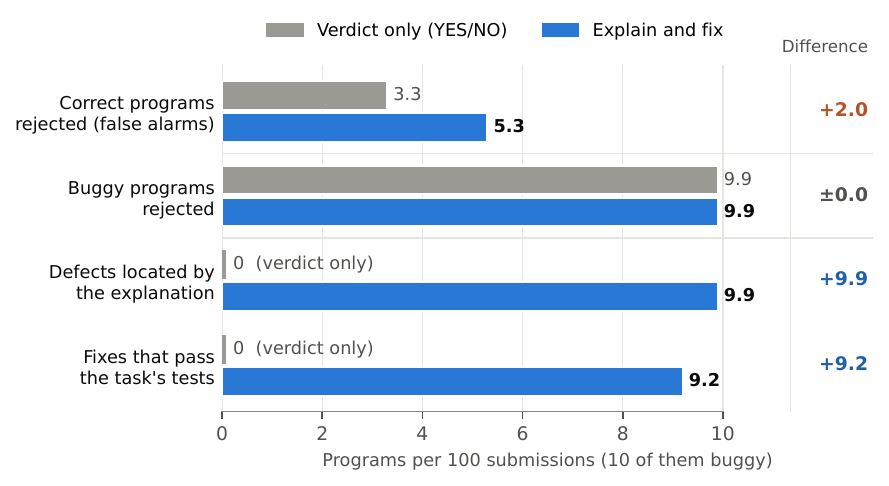}
\par\medskip
\begin{minipage}{\linewidth}
\textbf{Fig. 1} Expected review outcomes per 100 submissions with 10 buggy programs, using the adjudicated rates (Section 5.1). Both prompts flag nearly every buggy program. Explain-and-fix review flags about 2.0 additional correct programs. Among the buggy programs it flags, the explanation locates the defect in nearly every review (exploratory assessment by model judges), and most proposed fixes pass the task's tests
\end{minipage}
\end{figure}

How often does a review deliver that much, and how many correct programs does it flag along the way? On three paired benchmarks, Jin and Chen (2026) found that asking language models to explain a rejection and propose a fix increased false rejections. Their fix-guided filter used tests to reverse many of them. Measuring that cost requires a trustworthy reference. The released Mostly Basic Programming Problems (MBPP) tests define their assertions without calling them, and 533 MBPP pairs represent 91 distinct requirements. An incorrect label on a recurring program can therefore count against the same reviewer several times.

We used the original programs, prompts and filter to measure how often a richer review changes the verdict and what the proposed edit does when executed. Six models from six vendors reviewed the same programs under each prompt. We audited the released Python programs and the Java translations of the HumanEval tasks against the strongest available public tests, then checked disputed labels against the written requirements. We also varied the execution evidence available to the reviewer. Five questions guided the replication.

\begin{itemize}
\item \textbf{RQ1.} How many released programs carry a label that an independent test oracle does not confirm, and how many distinct tasks stand behind each benchmark's counts?
\item \textbf{RQ2.} On the audited programs, how often do current models reject correct code, and does the rate rise from a verdict-only prompt to the prompt that also asks for an explanation and a fix?
\item \textbf{RQ3.} When a model rejects correct code and proposes a fix, what does the fix do on the independent tests?
\item \textbf{RQ4.} Does giving the reviewer the results of running the requirement's examples change false rejection and false acceptance?
\item \textbf{RQ5.} Does the fix-guided filter still reduce false rejection when it verifies with tests a reviewer would have, and with the MBPP tests actually run?
\end{itemize}

The two prompts differed little in how often they accepted buggy programs. With adjudicated labels, the false-acceptance rates were 1.0\% for a verdict alone and 0.8\% for a verdict with an explanation and fix. The difference for a developer came after a correct rejection. Among correct rejections of audited buggy programs, 99.5\% of explanations identified the planted defect according to the exploratory model assessment. Proposed edits passed the task's tests in 93.1\% of those reviews. On the subset assessed with the strongest suites, HumanEval+ and MBPP+, the pass rate was 85.1\%. Figure 1 relates these outcomes to the additional false alarms. Section 5.1 gives the comparison at two assumed defect rates.

The replication makes four contributions.

\begin{enumerate}
\item We audit the paired benchmarks, then test counterexamples for every audited canonical program rejected under the explain-and-fix prompt. The replication package records each disputed label and the reason for each exclusion.
\item We estimate the paired prompt effect in two languages, under two verdict parsers, against a practical margin specified before the main run.
\item We execute proposed fixes for correct and buggy programs and measure the effect of showing reviewers the outputs of the requirement's examples. A separate exploratory assessment checks whether explanations locate the planted defect.
\item We rerun the original filter with the released tests, the tests with their assertions called, and the public examples a reviewer could consult.
\end{enumerate}

\section{Related work}

\subsection{Language models as judges of code}

Zheng et al. (2023) examine position, verbosity and self-enhancement biases of large language model (LLM) judges, and a recent survey treats the reliability of such judges as an open problem (Gu et al. 2024). For code, Tong and Zhang (2024) found that an LLM evaluator's demands for unnecessary error handling made it too strict on partially correct code, one of the failure patterns they identified. Crupi et al. (2025) found that LLM judges underestimate the correctness of human-written code in Java and Python, and Moon et al. (2025) showed that code judges respond to surface features that should not change a correctness verdict. Thinking models judge code better than non-thinking ones on CodeJudgeBench (Jiang et al. 2025), which is one reason to re-measure an effect found with models of an earlier generation. The closest current-model study is RTLReviewBench (Mahmud 2026). Twelve current commercial and open-weight models reviewed matched pairs of correct and buggy Verilog modules with labels confirmed by simulation; they rejected between 2\% and 59\% of correct modules, while all but one missed at most 3.5\% of buggy ones. Our paired design extends this question to Python and Java and varies what reviewers must return with their verdict.

Jin and Chen (2026) and Jin and Chen (2025) ask whether a program satisfies a written requirement. Reusing their paired programs lets us score the same model answers twice, first against the released labels and then against audited labels. This comparison isolates the contribution of benchmark scoring within our replication.

\subsection{Self-correction, critique and the harm of edits}

Self-Refine improved outputs through repeated model feedback (Madaan et al. 2023), and Self-Debugging lets a model investigate execution results (Chen et al. 2023). The quality of the feedback limits self-repair (Olausson et al. 2023). Without external feedback, self-correction can degrade an initially correct answer (Huang et al. 2023). Reliable feedback from execution can make self-correction effective (Kamoi et al. 2024). Longer critiques can also invent more defects or dwell on trivial issues (McAleese et al. 2024). For code already passing its tests, both self-refinement without execution feedback (Kim 2026) and a reviewing agent's edits (Xiang et al. 2026) can introduce failures. Across six task domains, including coding, agents damaged work verified as correct when told it was faulty (Mao et al. 2026). Cihan et al. (2025b) measured how often review suggestions worsen canonical HumanEval solutions. Assistants' tendency to agree with the premise of a request (Sharma et al. 2023) offers one possible explanation for editing correct code when asked to find and fix a defect. RQ3 tests what happens to correct programs when reviewers reject and edit them under Jin and Chen's Full prompt.

\subsection{Execution as evidence for a reviewer}

CodeT selects candidate programs by running them on generated tests (Chen et al. 2022). Execution-free critics instead predict test outcomes (Yadavally et al. 2025). The cost of building test environments motivated the use of code judges and, in turn, benchmarks for evaluating those judges (Wang et al. 2025). In an upper-bound diagnostic, five of six reviewers rejected fewer acceptable patches from a coding agent when given official test results in place of unchecked structured evidence. Two then judged every patch correctly (Guo et al. 2026). Simply telling two LLM judges that all tests passed had no significant effect on their judgments in another study (Zhan et al. 2026). A model's own tests can also reject correct programs (Chen et al. 2025), a concern for the fix-guided filter. RQ4 examines a source available to the reviewer: the results of running examples in the requirement. We measure both false rejection and false acceptance because the evidence could reduce one at the expense of the other.

\subsection{Labels, oracles and benchmark errors}

Label errors in benchmark test sets are common and can change which model looks best (Northcutt et al. 2021). Annotation errors distort the performance and rankings reported on text-to-SQL benchmarks (Jin et al. 2026), and many patches that pass SWE-bench tests behave differently from the reference (Wang et al. 2026a). For code generation, EvalPlus extended HumanEval's tests with many generated inputs and found solutions that the original tests had accepted wrongly (Liu et al. 2023). The paired benchmarks of Jin and Chen derive from HumanEval (Chen et al. 2021), MBPP (Austin et al. 2021) through a bug-aware corpus (Wang et al. 2024), QuixBugs (Lin et al. 2017; Ye et al. 2021) and HumanEvalPack (Muennighoff et al. 2023). The original's own validity section notes that the paired design assumes the canonical solutions and tests reflect the requirement (Jin and Chen 2026). RQ1 tests that assumption.

\subsection{Re-tests of the overcorrection result and of the filter}

CIVER evaluates pull-request review against a baseline inspired by the fix-guided filter (Song et al. 2026). Our search found no independent re-test of the original filter implementation under different test oracles (Section 3.7). We follow the reporting guidance and terminology summarised by Carver et al. (2014). Our replication is external and differentiated: we worked independently, without contacting the original authors, and changed the study context and measurements. From their public repository we reused the programs, prompt templates, original verdict parser and filter code. Section 3 describes the new reviewers, label audit and additional outcomes.

\subsection{Language models in code review practice}

Developers use code review to understand a change as well as to find defects (Bacchelli and Bird 2013). Pre-trained models can automate parts of review (Tufano et al. 2022; Li et al. 2022). Industrial reviewers have produced useful findings alongside faulty reviews and unnecessary edits (Cihan et al. 2025a), and developers rejected most agentic review comments in one open-source study (Lin et al. 2026). These findings motivate measuring the quality of the diagnosis and proposed edit alongside the accuracy of the verdict.

\section{Method}

We follow the general ACM SIGSOFT Empirical Standard (Ralph et al. 2020) and the standards for replications, experiments and repository mining in the standards repository. Before the main run, we froze the design, analysis scripts and simulation. The replication package records the date and reason for each subsequent change.

\subsection{Items and the label audit}

\textbf{Released items.} We cloned the authors' repository at commit 642b2bf. It contains 164 HumanEval, 533 MBPP and 40 QuixBugs pairs. Each pair comprises a requirement, a canonical program, a buggy program and tests. Recurring MBPP requirements share a canonical program. All but 5 pairs have a distinct buggy variant for their requirement. We retained each distinct program once and recorded its multiplicity in the release.

\textbf{Oracles.} Each program was run against the strongest public tests for its task:
\begin{itemize}
\item HumanEval: HumanEval+ v0.1.10 (Liu et al. 2023), the task's base and added inputs, with expected outputs from the EvalPlus reference and its tolerance rules.
\item MBPP: MBPP+ v0.2.0 where it contains the task. We used an MBPP+ record only if its function took the released number of parameters and its reference passed the released assertions. For other tasks we invoked the released assertions as the oracle.
\item QuixBugs: the repository's own pytest suites.
\item Java: the HumanEvalPack test class.
\end{itemize}

A canonical program that passes its oracle and a buggy program that fails it have labels consistent with the test outcome. We excluded programs when the label disagreed with the oracle, a buggy program failed to compile, or the oracle could not assess the program. We treated a canonical timeout, an item that failed to load and a test that raised an exception before calling the program as unassessable. A timeout on a buggy program counted as a failure. Counterexamples also disqualified three canonical MBPP programs that passed the oracle. In one case the MBPP+ reference shared the defect. In the other two, the released assertions never exercised the failing input. The replication package lists the counterexamples. Section 4.1 describes the subsequent adjudication against the written requirements.

\subsection{Reviewers}

Six models from six vendors served as confirmatory reviewers: GPT-6 Luna, DeepSeek V4.1 Flash, Qwen 3.8 Flash, MiMo V2.6 Flash, GLM 5.3 Flash and Claude Haiku 5.5. We retained the original persona line in the user message. The original study set temperature to zero. We left temperature at the provider default (the original study used zero): the gateway route used for five reviewers passes no sampling settings, and the batch requests for Claude Haiku 5.5 set none.

Five models ran through a gateway with default reasoning settings, a short environment block and a one-line message prefix. We replaced the default agent's coding instructions with the system prompt "You are a helpful assistant." and disabled its tools. We accessed Haiku directly through its vendor's batch interface to use the access available there. It ran at high effort with an 8,000-token output limit and no agent harness.

When gateway credit ran out, the remaining 967 GPT-6 Luna answers (10 on audited items) came through the vendor's command-line agent in a read-only sandbox at high reasoning effort. The other confirmatory routes provided no tools. We recorded the route of each answer. Two models added after the design freeze, Claude Opus 5.5 and Qwen 3.7 Plus, form a separate extension (Section 4.8). During the rerun of 389 items with corrected execution evidence (Section 3.3), gateway credit again ran out. Subsequent Claude requests used the vendor's command-line client with the same system prompt, an added environment block, high effort and no tools. Subsequent GPT-6 Luna requests used the agent route described above. The answers were collected on 26 November 2026.

\subsection{Conditions}

Every audited program was judged by every reviewer under four conditions, in a seeded random order per reviewer:
\begin{itemize}
\item \textbf{Direct, Direct+Explain, Full:} the authors' three prompts, built by their own code, unchanged except for the system prompt and the route additions above. For Java the words "Python function" become "Java function" and the code fences follow.
\item \textbf{Full with execution evidence:} the Full prompt with one block inserted before its first step: the requirement's examples, run on the program, each with the value returned, the value expected and whether it matched.
\end{itemize}

We showed only calls printed in the requirement. For each call we used the benchmark's expected value if its tests included that call. Otherwise, we used the reference program's output only when it matched the value printed in the requirement. For 2 HumanEval problems and their Java translations, the printed example conflicted with the benchmark. Both the evidence block and the public examples used the benchmark's value. Sections 4.4 and 4.5 report sensitivity to excluding those 4 tasks.

The first internal review identified errors in the evidence delivered to reviewers. Checking every audited item revealed discrepancies for 389 of 1,036 programs, including 320 Java programs. Java items received a single aggregate line. Other discrepancies included calls absent from the requirement, missed examples and incorrect assessments. We corrected the extraction and repeated the affected requests. RQ4 uses those responses. To reproduce the original scoring, we also obtained verdicts on the released programs outside the audited set under the three original prompts.

\subsection{Measures}

\textbf{Verdict.} After stripping Markdown formatting and step labels, the strict parser accepts the first line consisting of YES or NO, or an explicit "Answer: YES/NO" line. It records a missing verdict if neither occurs. The original parser takes the first YES or NO anywhere in the response. A model from another family read 160 pilot answers blind to the parser's verdict and agreed with the strict parser on all of them. Validation used no human readers. For the 28 main-run answers on which the parsers disagreed (Section 4.2), a model read each answer before seeing either verdict. It agreed with the strict parser in 13 cases and the original parser in 11, and judged 4 unclear. Most cases favouring the original parser began with "NO." followed by text on the same line, a form the strict parser leaves unclassified.

\textbf{Outcomes.} A NO on a correct program is a false rejection. A YES on a buggy program is a false acceptance. Each rate uses the programs of the corresponding class for which the parser returned a verdict.

\textbf{Fix outcome (RQ3).} For each Full response rejecting a program labelled correct, we extracted the last block defining code, skipping transcript blocks. If necessary, we used the block under the fix step. Following the original implementation, we assembled the fix with the item's declaration and ran it on the same oracle. We recorded whether the edit passed, failed the tests or could not run.

\textbf{Filter (RQ5).} We applied the authors' unchanged \texttt{apply\_\allowbreak{}filter} function, with their reversal rule enabled, to every Full-prompt NO with a fix. The filter takes the last code block in the response. For 5 answers this differed from the RQ3 extraction. Substituting the RQ3 fix changed 10 filter decisions. Under the hidden tests the filter still reversed 62 of 195 false rejections, with 1 reversal reclassified from a failing fix to equivalent behaviour. GPT-6 Luna replaced GPT-4o as the input generator. As in the original implementation, the generator saw the task's released example test. We generated inputs once per comparison and reused them across the three frozen arms.

The original execution helpers include process start-up in the time limit and can read a result before it arrives. Under load on our Windows machine, these behaviours made many correct programs fail their own tests. We reran the filter using the authors' unchanged worker functions inside a wrapper that times execution itself. We used three test oracles: the released tests as originally executed (MBPP assertions uncalled), the same tests with the assertions called, and the public examples described in Section 3.3.

In the public-example arm, the input generator still saw the released example test. After the first internal review we added a fourth arm in which the requirement's public examples served both as the oracle and as the generator's only examples. The benchmark values for the 4 tasks in Section 3.3 and one numerical tolerance were the only information beyond the requirement available to this arm.

\textbf{Explanation quality (exploratory).} For each Full-prompt review that correctly rejected an audited buggy program, a judge saw the requirement, the buggy and canonical programs, their line difference, and the reviewer's explanation with its fix code removed. The judge scored location (located, partly located, wrong defect or not mentioned) and whether the explanation correctly described a consequence. DeepSeek V4 Pro judged reviews from other vendors. LongCat 2.5 judged reviews written by DeepSeek V4.1 Flash.

We specified the measure after internal review and before judging any explanations. We subsequently assigned the DeepSeek reviews to the second vendor and added a harder control set before completing the judgments. Both judges classified all 60 initial controls correctly, distinguishing explanations based on the actual change from explanations taken from other tasks. On the harder controls they correctly classified 30 of 30 genuine explanations and 30 of 30 explanations rewritten to blame an unchanged line in the same program.

\subsection{Analysis}

For RQ2 we paired verdicts from the same reviewer on the same program, calculated the difference in false-rejection rate between Full and Direct for each reviewer, and averaged the differences with equal weights. RQ4 compares Full with execution evidence to Full alone. RQ5 compares false rejection after filtering with public examples to false rejection after filtering with hidden tests. We computed intervals over tasks, grouping each HumanEval problem with its Java translation. Inference is conditional on the six confirmatory reviewers.

Before the main run we specified five percentage points as the smallest change in false rejection that would warrant switching prompts. This margin is generous relative to our Direct rate, where five points is about half the baseline. The original study's Table 2 gives increases of +3.8 to +52.0 points (our average, 18.8). Our frozen method misquoted this range as 10 to 47 points. For RQ2 and RQ4 we tested for a change in either direction and used two one-sided tests to assess equivalence within five points in either direction (Lakens 2017). For RQ5 the frozen plan treated a null result as inconclusive. We added the equivalence test after the freeze and identify it as such.

We report cluster-robust 95\% confidence intervals (CIs), centring each pair's contribution at its reviewer's estimate (Cameron and Miller 2015). These intervals assume a sufficient number of independent task clusters. Correct programs fall in 281 clusters. Resampling task clusters for the RQ2 and RQ4 differences gave intervals within a few tenths of a percentage point. For a robustness check, we fitted a logistic model by generalised estimating equations (GEE) with task clustering, exchangeable within-cluster correlation and a condition-by-reviewer term (Liang and Zeger 1986). We use Wilson intervals for the coding agent's rates. RQ1 covers the full release, so its exact Clopper-Pearson intervals are descriptive (Clopper and Pearson 1934). RQ3 reports a ratio over tasks.

Before the main run we simulated the planned data structure 200 times. The analysis detected a five-point difference in every run. Its 95\% intervals covered the true value in 95.5\% of runs for RQ2 and RQ4 and in 91.5\% for RQ5. The simulation used a fixed parameter for correlation between arms and approximate groups for translation clusters. In particular, it did not reproduce the shared pre-filter verdict in RQ5. These runs check the estimators under the simulated conditions and offer limited evidence about coverage in the study itself.

The intervals treat the task clusters as draws from a population of similar short programming tasks with written requirements. Our sample consists of the audited release and the Java translations, and the six reviewers are fixed. In a separate exploratory comparison, OpenAI's Codex command-line agent reviewed the audited HumanEval programs under Full. Session logs identified its default model as GPT-6.2 Astra at medium reasoning effort. It had permission to run Python. When the sandbox blocked the interpreter, the agent's automatic approval allowed execution outside the sandbox.

\subsection{Pilot}

Before the main run, all six reviewers completed the full procedure on a seeded 5\% sample of audited tasks. A model from another family reviewed the pilot. An incorrect function name made the execution evidence label every HumanEval program unloadable. Other problems included assertions copied from benchmark tests, MBPP+ records matched to the wrong tasks and the three defective canonical programs described in Section 3.1. We corrected each problem and reran the pilot before starting the main study.

\subsection{Novelty search}

We searched OpenAlex, arXiv and Crossref for work from 2025 onwards, using eight queries and five patterns intended to refute the novelty claim. We read the closest matches in full. Two independent prior-work checks found no work that re-runs the authors' filter code or compares the same reviewer with and without results from the requirement's own examples on paired correct and buggy functions; the database search's only matches were the original study and its preprints.

\section{Results}

\subsection{Label audit (RQ1)}

\textbf{The released MBPP tests are never run.} Each test string defines \texttt{check(candidate)} with three assertions but never calls the function. The released verification step therefore passed every MBPP function that loaded, including 533 of 533 canonical programs and 252 buggy programs. Calling the assertions reduced those counts to 507 canonical programs and 0 buggy programs. HumanEval and QuixBugs do invoke their tests. Even so, 12 of 164 buggy HumanEval programs and 35 of 40 buggy QuixBugs programs passed the released tests.

\textbf{Labels against the independent oracles.} The 533 released MBPP pairs cover 91 requirements and 91 distinct canonical programs (Table 1). Each canonical program thus appears in about six pairs on average. Of 528 distinct buggy variants, 277 failed to compile. Most failures were indentation errors, which the original prompts instruct reviewers to ignore. We excluded the non-compiling variants because the oracle could not execute them to check their labels.

Across the 1,027 distinct released programs, the label disagreed with the oracle or with a known counterexample for 3.3\% (95\% CI 2.3\% to 4.6\%), the program did not compile for 27.2\% (24.5\% to 30.0\%), and the oracle could not assess 0.6\% (0.2\% to 1.3\%). The disagreements are concentrated in canonical programs: 17 HumanEval and 8 MBPP canonical programs fail the stronger tests, and the MBPP ones are counted 46 times in the released pairs. The authors of HumanEval+ had already reported that HumanEval's ground-truth solutions can be erroneous (Liu et al. 2023). For MBPP, an MBPP+ record was used for 31 tasks; 4 records were refused because their reference failed the released assertions or took a different number of parameters, and 56 tasks have no MBPP+ record.

\begingroup\footnotesize\setlength{\tabcolsep}{3pt}
\setlength{\LTpre}{\medskipamount}\setlength{\LTpost}{\medskipamount}
\begin{longtable}{>{\raggedright\arraybackslash\nobreak\hspace{0pt}}p{\dimexpr 0.1802\linewidth-2\tabcolsep\relax}>{\raggedright\arraybackslash\nobreak\hspace{0pt}}p{\dimexpr 0.1780\linewidth-2\tabcolsep\relax}>{\raggedleft\arraybackslash\nobreak\hspace{0pt}}p{\dimexpr 0.0778\linewidth-2\tabcolsep\relax}>{\raggedleft\arraybackslash\nobreak\hspace{0pt}}p{\dimexpr 0.0778\linewidth-2\tabcolsep\relax}>{\raggedleft\arraybackslash\nobreak\hspace{0pt}}p{\dimexpr 0.0984\linewidth-2\tabcolsep\relax}>{\raggedleft\arraybackslash\nobreak\hspace{0pt}}p{\dimexpr 0.0915\linewidth-2\tabcolsep\relax}>{\raggedleft\arraybackslash\nobreak\hspace{0pt}}p{\dimexpr 0.0881\linewidth-2\tabcolsep\relax}>{\raggedleft\arraybackslash\nobreak\hspace{0pt}}p{\dimexpr 0.1293\linewidth-2\tabcolsep\relax}>{\raggedleft\arraybackslash\nobreak\hspace{0pt}}p{\dimexpr 0.0737\linewidth-2\tabcolsep\relax}}
\multicolumn{9}{@{}p{\linewidth}@{}}{\normalsize\raggedright \textbf{Table 1} The label audit. Prog.: distinct programs; Pairs: the programs counted once per released pair in which they appear; Disagr.: the label disagrees with the oracle; No comp.: does not compile; Unass.: the oracle cannot assess the program; Counterex.: a counterexample shows the program wrong; Aud.: audited.}\\[4pt]
\toprule
Benchmark & Variant & Prog. & Pairs & Disagr. & No comp. & Unass. & Counterex. & Aud. \\
\midrule
\endfirsthead
\multicolumn{9}{@{}l}{\emph{(continued)}}\\
\toprule
Benchmark & Variant & Prog. & Pairs & Disagr. & No comp. & Unass. & Counterex. & Aud. \\
\midrule
\endhead
\midrule
\multicolumn{9}{r@{}}{\emph{continued on the next page}}\\
\endfoot
\bottomrule
\endlastfoot
HumanEval & canonical & 164 & 164 & 17 & 0 & 4 & 0 & 143 \\
HumanEval & buggy & 164 & 164 & 5 & 2 & 0 & 0 & 157 \\
MBPP & canonical & 91 & 533 & 8 & 0 & 1 & 3 & 79 \\
MBPP & buggy & 528 & 533 & 0 & 277 & 0 & 0 & 251 \\
QuixBugs & canonical & 40 & 40 & 1 & 0 & 1 & 0 & 38 \\
QuixBugs & buggy & 40 & 40 & 0 & 0 & 0 & 0 & 40 \\
Java & canonical & 164 & 164 & 0 & 0 & 0 & 0 & 164 \\
Java & buggy & 164 & 164 & 0 & 0 & 0 & 0 & 164 \\
\end{longtable}
\endgroup

\textbf{Adjudication after the first internal review.} We ran the programs against their written requirements to investigate every label disagreement, every counterexample and every program labelled buggy that passed its oracle. The case records are in audit/label\_adjudication.yaml in the replication package. Among programs the audit set aside, we confirmed 19 defects, including 1 identified on later examination. For 8 programs the requirement left the input or expected answer open, and for 3 the oracle expected the wrong output. A further 4 programs labelled buggy met their requirements. We also confirmed a defect in 1 canonical program retained by the audit (HumanEval/141).

Among the non-compiling MBPP variants, 40 differed from their canonical program only in indentation. Under the prompt's instruction to ignore indentation, 28 reproduced a canonical program that passed the oracle and had no recorded defect. The benchmark therefore labelled these programs buggy even though its own criteria supported acceptance. Another 10 reproduced a defect in the canonical program, and the requirements left 2 cases unresolved. The audit counted all of these variants as non-compiling.

Replaying the filter's generated inputs (Section 4.5), and running the same inputs on the Java translations, showed 36 audited canonical programs failing a valid input and 5 failing an input whose validity the requirement leaves open. A later re-examination found 2 more audited canonical programs defective.

We then examined all 135 audited canonical programs rejected by at least one reviewer under Full, including 38 already shown to be defective. We ran counterexamples from the reviewers and from the requirements. This check found 39 additional defective programs, including 28 Java translations, many with language-specific faults such as integer overflow or negative remainders. In 22 cases the only defect appeared at a numeric limit. The remaining outcomes were 29 ambiguous cases, 23 with no defect found and 6 that retained their earlier classification. A final check of 27 programs rejected only under other prompts found 1 further defect and 7 ambiguous cases. A rejecting reviewer had named the defect in 29 of the 39 newly confirmed Full-prompt cases. Because the search concentrated on rejected programs, we use the audited labels for the primary analysis and the adjudicated labels for sensitivity analysis (Section 4.2).

The audited set holds 1,036 programs (424 correct, 612 buggy) from 450 tasks in 287 clusters. All results below are on this set unless stated otherwise.

\subsection{False rejection under the three prompts (RQ2)}

Across the six reviewers, the pooled false-rejection rate on audited programs rose from 10.6\% under Direct (95\% CI 8.3\% to 12.9\%) to 14.0\% under Direct+Explain and 16.0\% under Full (13.0\% to 19.0\%). Table 2 gives the rates by reviewer and condition. False acceptance remained low across prompts, at 1.0\% under Direct and 0.8\% under Full.

\begingroup\footnotesize\setlength{\tabcolsep}{3pt}
\setlength{\LTpre}{\medskipamount}\setlength{\LTpost}{\medskipamount}
\begin{longtable}{>{\raggedright\arraybackslash\nobreak\hspace{0pt}}p{\dimexpr 0.2846\linewidth-2\tabcolsep\relax}>{\raggedleft\arraybackslash\nobreak\hspace{0pt}}p{\dimexpr 0.0881\linewidth-2\tabcolsep\relax}>{\raggedleft\arraybackslash\nobreak\hspace{0pt}}p{\dimexpr 0.1808\linewidth-2\tabcolsep\relax}>{\raggedleft\arraybackslash\nobreak\hspace{0pt}}p{\dimexpr 0.0810\linewidth-2\tabcolsep\relax}>{\raggedleft\arraybackslash\nobreak\hspace{0pt}}p{\dimexpr 0.1238\linewidth-2\tabcolsep\relax}>{\raggedleft\arraybackslash\nobreak\hspace{0pt}}p{\dimexpr 0.2367\linewidth-2\tabcolsep\relax}}
\multicolumn{6}{@{}p{\linewidth}@{}}{\normalsize\raggedright \textbf{Table 2} False-rejection rate (share of correct programs judged NO) by reviewer and condition, strict parser, audited set. The last column is the paired difference Full minus Direct in percentage points, computed on the programs with a verdict under both prompts, so it can differ slightly from the difference of the two rates beside it. In the pooled row, rates use all verdicts and the paired difference weights the six reviewers equally. The lower block averages their rates within each benchmark.}\\[4pt]
\toprule
Reviewer & Direct & Direct+Explain & Full & Full with evidence & Full minus Direct (95\% CI) \\
\midrule
\endfirsthead
\multicolumn{6}{@{}l}{\emph{(continued)}}\\
\toprule
Reviewer & Direct & Direct+Explain & Full & Full with evidence & Full minus Direct (95\% CI) \\
\midrule
\endhead
\midrule
\multicolumn{6}{r@{}}{\emph{continued on the next page}}\\
\endfoot
\bottomrule
\endlastfoot
GPT-6 Luna & 24.8\% & 27.7\% & 27.4\% & 27.8\% & +2.6 (0.0 to 5.2) \\
DeepSeek V4.1 Flash & 8.3\% & 13.9\% & 14.6\% & 12.5\% & +6.4 (4.0 to 8.7) \\
Qwen 3.8 Flash & 10.1\% & 14.6\% & 18.4\% & 16.1\% & +8.3 (5.4 to 11.1) \\
MiMo V2.6 Flash & 7.6\% & 8.3\% & 11.7\% & 9.2\% & +3.8 (1.4 to 6.3) \\
GLM 5.3 Flash & 1.9\% & 3.5\% & 5.2\% & 3.8\% & +3.3 (1.2 to 5.4) \\
Claude Haiku 5.5 & 10.8\% & 16.3\% & 18.6\% & 17.2\% & +7.8 (4.9 to 10.6) \\
Pooled & 10.6\% & 14.0\% & 16.0\% & 14.4\% & +5.4 (4.0 to 6.7) \\
\midrule
\multicolumn{6}{@{}l}{\emph{Mean by benchmark}} \\
\addlinespace[2pt]
\hspace{1em}HumanEval & 5.5\% & 7.8\% & 8.8\% & 7.1\% & +3.3 (1.3 to 5.2) \\
\hspace{1em}MBPP & 11.8\% & 14.6\% & 18.0\% & 15.6\% & +5.9 (2.7 to 9.2) \\
\hspace{1em}QuixBugs & 10.1\% & 12.4\% & 15.4\% & 14.2\% & +5.3 (0.6 to 9.9) \\
\hspace{1em}Java & 14.5\% & 19.6\% & 21.5\% & 20.3\% & +6.9 (4.8 to 9.0) \\
\end{longtable}
\endgroup

\textbf{The prompt effect.} The Full prompt raised false rejection by 5.4 percentage points over Direct (95\% CI 4.0 to 6.7; 2,537 pairs in 281 clusters). The rise is different from zero, and its interval contains the five-point margin, so the data neither establish nor rule out a change of that size. Reviewer-specific estimates ranged from +2.6 points for GPT-6 Luna to +8.3 points for Qwen 3.8 Flash. Its unadjusted 95\% interval lay above zero for all six reviewers (for GPT-6 Luna only just), and with the original parser for five (MiMo V2.6 Flash's rise under that parser was +2.4 points, $-$0.2 to 4.9). Verdicts changed in both directions. Of 2,537 paired verdicts on correct programs, 158 went from YES under Direct to NO under Full and 22 the other way, one sampled answer per condition. Direct+Explain lay between the two: +3.4 points over Direct (2.3 to 4.5), and Full a further +1.9 points over Direct+Explain (0.9 to 2.9). False acceptance moved by a fraction of a point ($-$0.3 points, $-$0.7 to 0.1).

The estimated increase remained positive across sensitivity checks. Leaving out one reviewer at a time gave +4.8 to +5.9 points. Counting missing verdicts as rejections gave +5.6 points, and counting them as acceptances gave +5.4. The estimate was +5.4 after excluding the 24 items whose outcomes the analysts had seen in an early feasibility check with one reviewer. Restriction to answers from one server route or the batch interface gave +5.7. The prespecified logistic GEE with a reviewer term gave a geometric mean reviewer odds ratio of 1.79 (1.54 to 2.10). Fitting a common odds ratio across reviewers gave 1.62 (1.46 to 1.80).

The adjudicated set contains 961 programs after removing programs shown to be defective from the correct set and reinstating those wrongly excluded by the oracle. Across 2,082 paired verdicts on correct programs, false rejection was 3.7\% under Direct and 5.9\% under Full. The paired increase was +2.2 percentage points (95\% CI 1.1 to 3.2, 90\% CI 1.3 to 3.0), equivalent at the five-point margin in this post hoc sensitivity analysis. Retaining the 29 programs whose sole defect occurs at a numeric limit (22 from the systematic check of rejected programs) gave rates of 5.0\% and 8.7\%, a paired increase of +3.7 points (2.4 to 5.0). Thus many rejections scored as false under the audited labels concerned genuinely defective programs. With the original parser the audited Full-minus-Direct difference was +5.1 points (3.8 to 6.5). The parsers disagreed on 28 of 30,597 answers (0.09\%).

\textbf{Comparison with the original study.} We averaged the rates in Table 2 of Jin and Chen (2026) across their five models and three Python benchmarks, giving equal weight to each model and benchmark. Under their scoring of the released items, false rejection averaged 41.9\% under Direct and 60.7\% under Full. The mean increase was +18.8 points, with model-by-benchmark increases ranging from +3.8 to +52.0. Table 3 compares the benchmark means.

Using the same released items, original parser and weighting, our reviewers rejected 14.9\% of programs labelled correct under Direct and 20.3\% under Full, about a third as often as the original reviewers. The paired increase was +5.4 points (3.4 to 7.4). As in the primary pooled analysis, the interval spans the five-point margin. Paired verdicts were complete in 14 of 18 reviewer-by-benchmark cells. Coverage was lowest for MiMo V2.6 Flash on MBPP (495 of 533 copies). The relative increases were closer across studies. The Full-to-Direct rate ratio was 1.37 with the original scoring, 1.44 with the strict parser and 1.51 on the audited items, compared with 1.45 in the original study. We did not estimate intervals for these ratios.

In the original study, false acceptance fell from 11.5\% under Direct to 7.9\% under Full as false rejection rose. With the same scoring, we observed a smaller decline in false acceptance, from 6.8\% to 5.6\% ($-$1.2 points, $-$2.2 to $-$0.2). Some Full-prompt rates for the same models differ between Tables 2 and 4 of the original study, limiting the precision of the comparison. We also scored our answers on the released items with the strict parser, pooling across programs and counting each canonical MBPP program once per released pair. This calculation gives greater weight to recurring MBPP programs, including some that fail the stronger tests. The MBPP rates rose, and the overall Full-minus-Direct difference was +7.7 points (4.8 to 10.6).

\begingroup\small
\setlength{\LTpre}{\medskipamount}\setlength{\LTpost}{\medskipamount}
\begin{longtable}{>{\raggedright\arraybackslash\nobreak\hspace{0pt}}p{\dimexpr 0.1813\linewidth-2\tabcolsep\relax}>{\raggedright\arraybackslash\nobreak\hspace{0pt}}p{\dimexpr 0.2611\linewidth-2\tabcolsep\relax}>{\raggedleft\arraybackslash\nobreak\hspace{0pt}}p{\dimexpr 0.1513\linewidth-2\tabcolsep\relax}>{\raggedleft\arraybackslash\nobreak\hspace{0pt}}p{\dimexpr 0.2152\linewidth-2\tabcolsep\relax}>{\raggedleft\arraybackslash\nobreak\hspace{0pt}}p{\dimexpr 0.1861\linewidth-2\tabcolsep\relax}}
\multicolumn{5}{@{}p{\linewidth}@{}}{\normalsize\raggedright \textbf{Table 3} False-rejection rates in the original study (mean of five models, Jin and Chen 2026, Table 2) and here (mean of six reviewers), by benchmark and prompt. "As released" scores this study's answers on the released items with copies counted. This study's columns use the strict parser and Python programs only; Section 4.2 gives the means computed exactly as the original's, with its parser.}\\[4pt]
\toprule
Benchmark & Prompt & Original & This study, as released & This study, audited \\
\midrule
\endfirsthead
\multicolumn{5}{@{}l}{\emph{(continued)}}\\
\toprule
Benchmark & Prompt & Original & This study, as released & This study, audited \\
\midrule
\endhead
\midrule
\multicolumn{5}{r@{}}{\emph{continued on the next page}}\\
\endfoot
\bottomrule
\endlastfoot
HumanEval & Direct & 34.2\% & 9.9\% & 5.5\% \\
HumanEval & Direct+Explain & 46.7\% & 12.4\% & 7.8\% \\
HumanEval & Full & 55.2\% & 14.3\% & 8.8\% \\
MBPP & Direct & 52.9\% & 20.2\% & 11.8\% \\
MBPP & Direct+Explain & 60.9\% & 24.4\% & 14.6\% \\
MBPP & Full & 70.5\% & 28.9\% & 18.0\% \\
QuixBugs & Direct & 38.5\% & 12.1\% & 10.1\% \\
QuixBugs & Direct+Explain & 45.5\% & 14.3\% & 12.4\% \\
QuixBugs & Full & 56.5\% & 17.6\% & 15.4\% \\
\end{longtable}
\endgroup

\subsection{How proposed fixes perform on correct and buggy code (RQ3)}

\textbf{Edits to programs labelled correct (RQ3).} Across 134 tasks in 114 clusters, 406 Full-prompt rejections of audited correct programs contained an extractable fix. Of those edits, 335 passed the oracle, 50 failed its tests (12.3\%) and 21 did not run (5.2\%). In 17 cases the edit changed the method name or return type. In 4 it imported an unavailable package. Under the prespecified definition of harm, 17.5\% failed or did not run (95\% CI 9.9\% to 25.0\%).

The share that failed or did not run was 28.0\% on HumanEval (75 fixes), 14.7\% on its Java translations (211), 11.8\% on MBPP (85) and 25.7\% on QuixBugs (35). Some MBPP tasks have just three test assertions. Excluding programs later shown to be defective gave a harm rate of 28.9\% (35 of 121 fixes on 57 programs, 95\% CI 14.1\% to 43.7\%). Under Full with execution evidence, 307 fixes attached to false rejections passed, 43 failed and 15 did not run. Another 2 rejections contained no fix.

\textbf{Repairs to buggy programs.} We ran each proposed fix from a correct Full-prompt rejection against the same task oracle that the audited buggy program had failed. Of 3,643 such reviews, 3,391 produced an edit that passed (93.1\%). The pass rates were 91.5\% on HumanEval+, 68.0\% on MBPP+, 98.0\% on the released MBPP assertions, 98.5\% on the Java tests and 90.7\% on QuixBugs. Across the two strongest suites, HumanEval+ and MBPP+, the rate was 85.1\% (1,291 reviews). The denominator counts reviews, so the same program can contribute a proposed fix from more than one reviewer.

\subsection{Execution evidence (RQ4)}

Showing the reviewer the results of the requirement's own examples changed false rejection under the Full prompt by $-$1.5 percentage points (95\% CI $-$2.7 to $-$0.4). The change is different from zero but inside the five-point margin (equivalent by the two one-sided tests). The analysis includes 2,536 pairs and uses the corrected answers from every reviewer for the 389 affected items.

On the 647 items whose evidence was delivered as described from the start, the change was $-$1.9 points ($-$3.3 to $-$0.5). The re-asked evidence answers of Claude Haiku 5.5 and some of GPT-6 Luna came through another client than their Full answers; on the 1,973 pairs answered by the server or batch routes alone, the change was $-$1.9 points ($-$3.1 to $-$0.7). Without the 4 tasks whose printed examples conflict with the benchmark it was $-$1.2 points ($-$2.2 to $-$0.2). For the 55 audited MBPP tasks whose labels rest on the released assertions (no usable MBPP+ record), the evidence showed reviewers the same three assertions used to assign labels. Excluding their 55 correct and 192 buggy programs gave a change in false rejection of by $-$1.4 points ($-$2.5 to $-$0.2) and in false acceptance of $-$0.5 ($-$0.9 to $-$0.1).

Across the audited set, false acceptance changed by $-$0.5 points ($-$0.8 to $-$0.1). With the evidence, 5 rejections of buggy programs became acceptances and 22 acceptances became rejections. On correct programs, 96 rejections became acceptances and 57 acceptances became rejections. By benchmark the change in false rejection ranged from $-$2.3 points on MBPP to $-$1.1 on the Java translations ($-$2.9 to 0.6); it was $-$1.6 on HumanEval and $-$1.3 on QuixBugs, where 14 of the 40 tasks have no example in the requirement.

The GEE odds ratio for false rejection was 0.85 (0.77 to 0.94). On the adjudicated labels, execution evidence changed false rejection by $-$0.4 points ($-$1.4 to 0.5). This estimate uses 2,062 pairs because reinstated programs had no evidence responses. The interval includes zero, as does the interval when programs with numeric-limit defects are retained ($-$0.9 points, $-$2.0 to 0.2). On the audited labels the pooled rate with evidence was 14.4\% (11.7\% to 17.2\%), above the rate for Direct without evidence.

\subsection{The fix-guided filter under different oracles (RQ5)}

Under the Full prompt the six reviewers rejected audited correct Python programs 195 times, each time with a fix the filter could run. Verifying with the hidden tests, the filter reversed 62 of these rejections (31.8\%): 39 because the fix gave the same outputs as the program on every generated input, and 23 because the fix failed where the program did not (20 on the tests, 3 by raising an error on a generated input). It kept the other rejections: in 82 the fix gave different outputs from the correct program on some generated input, in 35 the correct program raised an error or did not finish on a generated input on which the fix ran, and 16 were inconclusive.

Replaying the cases in which the correct program failed showed that most are inputs the requirement allows, on which the "correct" program is in fact wrong (Section 4.1). We also replayed all 82 cases with different outputs (28 programs): the differing input was one the requirement allows in 64, and on it the requirement supports the fix's output in 57 and the correct program's in 2. Such inputs are cases the examples do not show, for example integers beyond floating-point precision, non-ASCII text, negative or fractional arguments, and arrays with repeated values. The pooled false-rejection rate fell from 12.5\% to 8.6\%, a change of $-$4.0 points (95\% CI $-$5.6 to $-$2.4; Table 4).

For all 91 MBPP tasks, the hidden and public assertions are identical. The hidden check is empty for 34 of 40 QuixBugs tasks, so only a public example can change a decision on test evidence there. Hidden tests add checks on 162 HumanEval tasks. After filtering, the difference in false rejection with public examples versus hidden tests was $-$0.1 points (95\% CI $-$0.2 to 0.1, 90\% CI $-$0.2 to 0.0). The two one-sided tests added after the freeze classified this difference as equivalent at the five-point margin.

The reviewers rejected a correct program in 195 of 1,555 pairs, across 68 tasks. In 1 pair the hidden tests kept the rejection and the public examples reversed it. The count in the opposite direction was 0. In the discordant QuixBugs case, the public test required an exact value. The reviewer's fix returned a different value within the tolerance allowed by the requirement.

The overall effect of the hidden-test filter was $-$4.0 points (95\% CI $-$5.6 to $-$2.4). Across all four oracles, false rejection after filtering ranged from 8.5\% to 9.1\%. With one discordant pair and uncertain interval coverage (Section 3.5), the counts show most directly how often the choice of oracle changed a verdict. The public-minus-hidden difference ranged from $-$0.4 to +0.0 points by reviewer. On the HumanEval tasks with extra hidden checks, it was +0.0 points across 851 pairs in 142 clusters with a verdict. Every pair agreed, leaving no informative interval for this subset. Generated inputs for the QuixBugs graph tasks had the wrong format (Section 6).

Comparisons of the fix and program on shared generated inputs accounted for 39 of 62 hidden-test reversals. The tests themselves accounted for 20 reversals with hidden tests and 21 with public examples. The estimated difference in false acceptance was +0.0 points ($-$0.2 to 0.2).

In the three frozen arms the generator of test inputs saw the task's released example test, which can hold calls the requirement does not show. In the public-only arm, added after the first internal review, the generator saw only the requirement's own examples (with the exceptions in Section 3.4): false rejection after the filter differed from the hidden-test arm by +0.0 points ($-$0.6 to 0.6), and false acceptance by +0.0 points ($-$0.2 to 0.2). Without the tasks whose printed examples conflict with the benchmark the difference was +0.0 points ($-$0.6 to 0.6). The arms disagreed on 18 decisions: 9 rejections were reversed only with the hidden tests and 9 only in the public-only arm.

With the released oracle, the filter reversed 54 false rejections overall. The MBPP assertions were never called. On MBPP it reversed fewer false rejections (14.2\% remained, against 12.5\% with the hidden tests), because a fix cannot fail tests that are not run. It also turned 19 correct rejections of buggy programs into acceptances, 8 of them on MBPP, against 11 with the hidden tests, none on MBPP; false acceptance changed by +0.7 points from no filter (0.1 to 1.3). Without the reversal rule for a fix that fails the tests, the filter reversed only rejections whose fix behaved identically, and the released, hidden-test and public-example oracles each left 10.0\% of correct programs rejected (the public-only arm 10.3\%) ($-$2.5 points from no filter, $-$3.5 to $-$1.5).

\begingroup\footnotesize
\setlength{\LTpre}{\medskipamount}\setlength{\LTpost}{\medskipamount}
\begin{longtable}{>{\raggedright\arraybackslash\nobreak\hspace{0pt}}p{\dimexpr 0.2787\linewidth-2\tabcolsep\relax}>{\raggedleft\arraybackslash\nobreak\hspace{0pt}}p{\dimexpr 0.1596\linewidth-2\tabcolsep\relax}>{\raggedleft\arraybackslash\nobreak\hspace{0pt}}p{\dimexpr 0.1624\linewidth-2\tabcolsep\relax}>{\raggedleft\arraybackslash\nobreak\hspace{0pt}}p{\dimexpr 0.1454\linewidth-2\tabcolsep\relax}>{\raggedleft\arraybackslash\nobreak\hspace{0pt}}p{\dimexpr 0.1139\linewidth-2\tabcolsep\relax}>{\raggedleft\arraybackslash\nobreak\hspace{0pt}}p{\dimexpr 0.1351\linewidth-2\tabcolsep\relax}}
\multicolumn{6}{@{}p{\linewidth}@{}}{\normalsize\raggedright \textbf{Table 4} False rejection and false acceptance after the fix-guided filter, by oracle (Full prompt, six reviewers, audited correct and buggy Python programs, reversal rule on). The last three columns are false rejection by benchmark. Rates are on the Python programs with a Full-prompt verdict (1,555 of the 1,560 judgments of correct programs; the others had no verdict the strict parser could read), pooled over reviewers, while Table 2's benchmark rows average the six reviewers' rates, so the no-filter row can differ slightly from Table 2. Public only: the public examples as the oracle and as the generator's only examples. The original's rates are its five-model means before and after its filter (Jin and Chen 2026, Table 4).}\\[4pt]
\toprule
Oracle & False rejection & False acceptance & HumanEval & MBPP & QuixBugs \\
\midrule
\endfirsthead
\multicolumn{6}{@{}l}{\emph{(continued)}}\\
\toprule
Oracle & False rejection & False acceptance & HumanEval & MBPP & QuixBugs \\
\midrule
\endhead
\midrule
\multicolumn{6}{r@{}}{\emph{continued on the next page}}\\
\endfoot
\bottomrule
\endlastfoot
None & 12.5\% & 0.8\% & 8.8\% & 18.0\% & 15.5\% \\
Released tests & 9.1\% & 1.5\% & 6.1\% & 14.2\% & 9.7\% \\
Hidden tests & 8.6\% & 1.2\% & 6.1\% & 12.5\% & 9.7\% \\
Public examples & 8.5\% & 1.2\% & 6.1\% & 12.5\% & 9.3\% \\
Public only & 8.6\% & 1.2\% & 6.2\% & 12.7\% & 8.8\% \\
Original, before &  &  & 54.8\% & 69.0\% & 51.0\% \\
Original, after &  &  & 16.3\% & 28.9\% & 24.0\% \\
\end{longtable}
\endgroup

In the original, the filter cut false rejection under Full by more than half on every benchmark. Here it removed 62 of 195 false rejections (31.8\%) from a much smaller rate. On the rejections of programs the adjudication did not show defective it removed 30 of 58 on 29 programs (51.7\%), a sensitivity figure, limited by the programs left ambiguous and by an adjudication made after the results were known. In 117 of the 195 the fix and the correct program behaved differently on some generated input. (The exploratory count declared before the main run, taken under the public oracle over the comparisons that reached generated inputs, is 116 of 158.)

\subsection{Do explanations locate the planted defect? (exploratory)}

Across 249 task clusters, the Full prompt produced 3,643 correct rejections of audited buggy programs. In 99.5\% of those reviews (95\% CI 99.2\% to 99.7\%), the judge found that the explanation identified the planted change. The share ranged from 99.2\% to 99.8\% across reviewers and was about 99\% or more on each benchmark. Including partial identifications raised the share to 99.9\%. In 4 reviews the explanation blamed a different defect. Reviewers correctly described a consequence of the planted defect in 99.1\% of reviews.

Jin and Chen (2026) scored agreement with bug-type categories (38\% to 75\%) and found that reviewers described consequences more often than causes. Our judges assessed whether an explanation pointed to the specific change in the program. A reviewer can identify that change without using the benchmark's category name.

\subsection{A coding agent as reviewer (exploratory)}

The coding agent judged the 143 correct and 157 buggy audited HumanEval programs under the Full prompt, with permission to run code. It rejected 26 correct programs (18.2\%, 95\% CI 12.7\% to 25.3\%) and accepted 0 buggy ones (0.0\%, 0.0\% to 2.4\%). Its full session logs show commands in 171 of its 300 runs. In 168 of them it first replaced the program under review with its own fix and ran only that, in 2 every attempt to run a program failed, and in 1 it ran the program as given; every run with commands ended in NO. Its false-rejection rate on these programs was above the six reviewers' Full-prompt mean on HumanEval (8.8\%) and its interval includes the rate of GPT-6 Luna, its vendor's chat model, without tools (16.8\%). The comparison varies both tool access and model family.

\subsection{Stronger siblings (extension)}

Claude Opus, the stronger sibling of Claude Haiku in the same release, showed a Full-minus-Direct rise of +1.9 points (0.6 to 3.2) against Haiku's +7.8, and execution evidence changed its false rejection by +2.4 points (0.8 to 3.9). On the 237 pairs whose two answers came through the batch interface, where the client did not change with the condition, the change was +1.3 points ($-$0.2 to 2.7); on the other 187 pairs the evidence answer also came through another client.

Qwen 3.7 Plus, the stronger sibling of Qwen 3.8 Flash, showed a rise of +2.1 points ($-$0.4 to 4.6) against Flash's +8.3, and execution evidence changed its false rejection by +0.0 points ($-$2.0 to 2.0). In both pairs the stronger model rejected fewer correct programs under the Full prompt (3.8\% against 18.6\%, and 9.4\% against 18.4\%), and its rise from Direct to Full was smaller. These comparisons describe the two model pairs tested.

\section{Discussion}

\subsection{What a developer gets from an explain-and-fix review}

On the adjudicated labels, reviewers accepted 1.0\% of buggy programs under Direct and 0.8\% under Full. Both prompts thus caught nearly the same share of buggy programs. What Full added was evidence a developer could inspect. In the exploratory assessment, model judges found that the explanations identified the planted change in nearly every correct rejection (Section 4.6). Developers also received a proposed edit they could run against the task's tests (Section 4.3). Direct+Explain also supplies a rationale and had a lower false-rejection rate on audited programs (14.0\% versus 16.0\% under Full). We assessed explanation quality under Full alone. Table 5 gives the projected costs and outputs per 100 submissions.

\begingroup\small
\setlength{\LTpre}{\medskipamount}\setlength{\LTpost}{\medskipamount}
\begin{longtable}{>{\raggedright\arraybackslash\nobreak\hspace{0pt}}p{\dimexpr 0.5957\linewidth-2\tabcolsep\relax}>{\raggedleft\arraybackslash\nobreak\hspace{0pt}}p{\dimexpr 0.1997\linewidth-2\tabcolsep\relax}>{\raggedleft\arraybackslash\nobreak\hspace{0pt}}p{\dimexpr 0.1997\linewidth-2\tabcolsep\relax}}
\multicolumn{3}{@{}p{\linewidth}@{}}{\normalsize\raggedright \textbf{Table 5} Expected review outcomes per 100 submissions at two assumed defect rates. Verdict counts use the adjudicated rates (Section 4.2). Counts of located defects and passing fixes apply the conditional rates from audited buggy programs (Sections 4.6 and 4.3, respectively) to the expected correct rejections. Defect location was assessed by model judges in an exploratory analysis. ``None'' denotes an output the verdict-only prompt does not request.}\\[4pt]
\toprule
Per 100 submissions & Verdict only (Direct) & Explain and fix (Full) \\
\midrule
\endfirsthead
\multicolumn{3}{@{}l}{\emph{(continued)}}\\
\toprule
Per 100 submissions & Verdict only (Direct) & Explain and fix (Full) \\
\midrule
\endhead
\midrule
\multicolumn{3}{r@{}}{\emph{continued on the next page}}\\
\endfoot
\bottomrule
\endlastfoot
\multicolumn{3}{@{}l}{\emph{10 buggy programs}} \\
\addlinespace[2pt]
\hspace{1em}Correct programs rejected & 3.3 & 5.3 \\
\hspace{1em}Buggy programs rejected & 9.9 & 9.9 \\
\hspace{1em}Defects located by the explanation & none & 9.9 \\
\hspace{1em}Fixes that pass the task's tests & none & 9.2 \\
\midrule
\multicolumn{3}{@{}l}{\emph{30 buggy programs}} \\
\addlinespace[2pt]
\hspace{1em}Correct programs rejected & 2.6 & 4.1 \\
\hspace{1em}Buggy programs rejected & 29.7 & 29.8 \\
\hspace{1em}Defects located by the explanation & none & 29.6 \\
\hspace{1em}Fixes that pass the task's tests & none & 27.7 \\
\end{longtable}
\endgroup

With 10 buggy programs per 100 submissions, each prompt would catch about 9.9. At the adjudicated rates, asking for an explanation and fix would yield 9.2 edits that pass the task's tests at the cost of 2.0 additional false rejections. The cost depends on how programs with a defect only at a numeric limit are classified. Retaining them among the programs scored as correct raises the projected cost to 3.4 per 100 at 10\% defect prevalence and 2.6 at 30\%. At 10\% prevalence, the audited labels give 4.9 additional false rejections. A specific diagnosis and a proposed edit could shorten a developer's search for the fault. A timed study could weigh that saving against the time spent checking additional false alarms.

\subsection{Why the measured cost of a rejection changed}

All six reviewers reproduced the direction of the original prompt effect (Table 2). With the original scoring, the absolute increase in false rejection was 5.4 percentage points here and 18.8 in the original. Figure 2 follows the overall false-rejection rates from the original study to this one, step by step. The relative increases were closer: Full rejected 1.37 to 1.51 times as many correct programs as Direct across our scoring methods, compared with 1.45 in the original. These ratios are descriptive. The distinction matters to a team reviewing a fixed volume of code: the same proportional increase produces fewer extra reviews when the baseline rate is lower. The pooled effect also conceals changes in both directions. With one answer per prompt, 6.2\% of paired verdicts on audited correct programs changed from accept to reject, and 0.9\% changed from reject to accept.

\begin{figure}[!tbp]
\centering
\includegraphics[width=\linewidth,keepaspectratio]{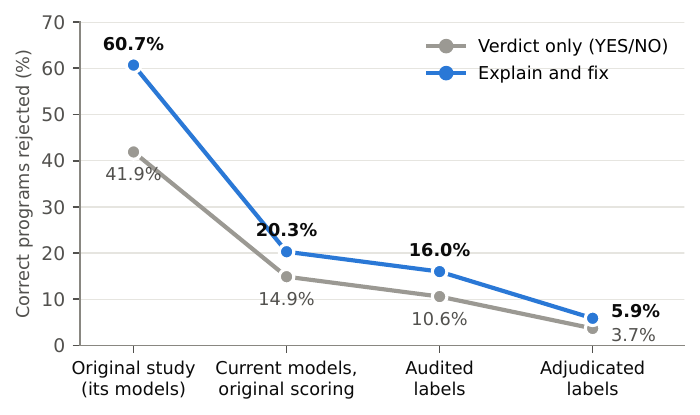}
\par\medskip
\begin{minipage}{\linewidth}
\textbf{Fig. 2} False rejection across four comparisons. The first two points use the original scoring on the released items: the original five models (our averages of their Table 2) and our six reviewers. Models, sampling and setup changed between these points. The third point uses audited labels and also adds Java programs, applies the stricter verdict parser and pools all verdicts. The fourth uses adjudicated labels. Its rates may slightly underestimate false rejection because the search for defects concentrated on rejected programs (Section 6)
\end{minipage}
\end{figure}

Scoring the released items inflates our measured rates and likely affected the original rates in the same direction. The gap between studies persists under matched scoring. Under the released scoring, 8 MBPP canonical programs that failed the stronger tests counted 46 times. Repeated requirements and defective canonical programs both affect the MBPP rates, and their individual contributions remain unquantified. Among canonical programs set aside by the audit, adjudication found 18 defective programs, 8 ambiguous requirements and 3 incorrect test expectations. On the buggy side, 28 non-compiling MBPP variants reproduce an apparently correct canonical program once indentation is restored. The prompt expressly tells reviewers to ignore indentation, yet accepting these variants counts against them.

Reviewers, sampling settings, system prompt, access routes and date changed together between studies, so their individual contributions to the gap remain unresolved. The original paper and its code give different identifiers for the Claude and Mistral models, leaving their precise versions uncertain.

\subsection{What execution reveals about a proposed edit}

Passing a test suite leaves a further question: where does the edit change behaviour? Of 195 rejections of audited correct Python programs, the filter found matching behaviour on all generated inputs in 39 cases and distinguished the two programs in 117. Another 16 comparisons were inconclusive, and in 23 the fix failed a test or raised an error where the original program ran. In the cases with two different outputs on a valid input, the requirement usually supported the fix (Section 4.5). An edit to a canonical solution can thus supply a counterexample to the benchmark label. It can also introduce a fault: among fixes to programs still classified as correct after adjudication, 28.9\% failed or did not run (95\% CI 14.1\% to 43.7\%).

Providing the results of the requirement's own examples reduced false rejection by 1.5 points on the audited labels. The interval on the adjudicated labels included zero (Section 4.4). This approach requires runnable examples with expected outputs, which were absent for 4 HumanEval and 14 QuixBugs tasks. Its cost and coverage in industrial requirements remain open. The exploratory agent study illustrates a different use of execution. The agent usually ran its own rewrite when it invoked a command. That run checks the edit and leaves the reported fault in the submitted program untested. A generated test can itself encode an incorrect expectation (Chen et al. 2025).

\subsection{Implications}

These implications are limited to short, self-contained functions, the six reviewers we tested at default sampling and the original's prompts (Section 6).

\textbf{For teams choosing a review prompt.} A verdict-only prompt incurs fewer false alarms when a pipeline consumes only the decision. A reviewer working with a developer can also supply the suspected fault and a candidate edit. At 10\% assumed defect prevalence, 34.7\% of Full-prompt rejections would concern correct code under the adjudicated rates (59.2\% under the audited labels). At 30\% prevalence those shares fall to 12.1\% and 27.3\%, respectively. This gives a developer a reason to check the reported fault against the requirement before blocking a change, and to test the proposed edit on the cited case before applying it. A second review sometimes changed the verdict. Of 167 initial rejections of audited correct programs sampled twice, 28 became acceptances. A two-step workflow that requests the diagnosis after a rejection would require a second call and remains untested, as do repeated-sampling gates and confidence thresholds. A gate on new code has the project's existing tests, but they may not cover the behaviour the change adds, so a fix-guided filter there may lack the oracle the benchmarks supply.

\textbf{For benchmark users.} Invoke the MBPP assertions and count distinct programs before scoring reviewers. Inspect non-compiling buggy variants separately, especially when the review instructions allow indentation errors. The replication package identifies every exclusion and records the adjudication for disputed programs.

\textbf{For researchers comparing reviewers.} A reproducible false alarm needs both the reviewer's objection and a defensible correctness label. Releasing the prompts, programs, scoring code and filter made this comparison possible. Recording the executable counterexamples alongside disputed labels would let later studies test the objection directly. Filter evaluations should also identify which inputs came from the requirement and which came from hidden tests. Here, many reversals followed from comparing the submitted program with the proposed fix on generated inputs shared by the three frozen filter arms (Section 4.5).

\section{Threats to validity}

\textbf{Construct validity.} Agreement between a program and its test oracle can preserve a defect they share. The MBPP+ reference shares a defect with at least one canonical program. For 60 of the 91 MBPP tasks, the strongest usable oracle consists of three released assertions. A requirement can also conflict with its own example. One MBPP task asks for the smallest natural number whose factorial is divisible by the input. Its example gives 4 for an input of 5, although 4! equals 24. The canonical program also returns 6 for an input of 7, which does not divide 6! evenly. It passed the audit and was removed after adjudication.

The first adjudication, the replay, re-examination and systematic check identified 79 audited canonical programs that fail on valid inputs. Direct checks concentrated on rejected programs. Transferring counterexamples reached one Java program that every Full reviewer had accepted, and checking programs rejected only under other prompts exposed another defect among 27. Undiscovered defects in accepted programs can inflate the denominator of the corrected false-rejection rate. We therefore treat adjudication as a sensitivity analysis and retain the audited set for the primary estimate.

The parsers extract verdicts, while the test suites evaluate observable behaviour. We checked parser agreement against model readings of the responses (Section 3.4). A passing edit can still change behaviour outside the tests. For explanations, one model judge assessed each answer against the planted change. Both judges passed the controls, including explanations that blamed an unchanged line in the same program. Agreement with human reviewers on these explanations remains unmeasured.

\textbf{Internal validity.} Each reviewer saw each program under each prompt, in a seeded random order. One answer per condition leaves sampling variability in the paired comparison. Where two runners overlapped, we retained the first answer to 3,341 requests answered twice. The verdict changed on the repeat in 88 cases (2.6\%). The gateway switched from one command-line process per call to server processes with the same message. It added the prefix "Reply to the message above." The Anthropic batch interface omitted that prefix. Within-reviewer comparisons hold these features constant when the route is shared. Sections 4.2 and 4.4 report sensitivity to the route changes described in Section 3.2, including the corrected execution-evidence requests.

A single model generated the filter inputs. The JSON representation gave the QuixBugs graph programs malformed inputs where their functions expected node objects or edge dictionaries keyed by node pairs. Most comparisons on those tasks ended inconclusively. On shortest\_path\_lengths the filter reversed four false rejections: twice because the fix and program agreed on malformed inputs and twice because the fix failed on them. One reversal on shortest\_paths rested on empty graphs. At least five reversals in Section 4.5 therefore depend on inputs that do not establish the intended graph behaviour.

\textbf{External validity.} The estimates apply to short, self-contained functions with written requirements, the six named reviewers at their November 2026 versions, and the prompt wording and sampling settings in Section 3. Industrial reviews involve larger changes, project context and non-functional requirements. The Java arm adds a language through translations of the HumanEval tasks, which share a cluster with their Python counterparts. The exploratory agent results describe one agent at one setting.

\textbf{Conclusion validity.} The intervals reflect variation across tasks for the six fixed reviewers. We grouped each HumanEval task with its Java translation. Before the main run, we specified the smallest effect of interest and the equivalence tests for RQ2 and RQ4. The RQ5 equivalence test and the public-only filter arm were later additions. The pre-run simulation omitted RQ5's shared pre-filter verdict, and its RQ5 intervals covered the true value in 91.5\% of runs, below the nominal 95\%. For RQ2 and RQ4, the task-cluster bootstrap gave intervals within 0.2 points of those reported here (replication package). We made the changes arising from pilot review before inspecting main-run outcomes. The first internal review took place after all results were available. The replication package documents the resulting corrections to the execution evidence, fix and filter harnesses, variance formula and prespecified GEE, gives the reason for each correction, and contains the scripts for every reported value.

\section{Conclusion}

Both prompts caught nearly the same share of buggy programs. What the developer received after a correct rejection differed. Under explain-and-fix review, the explanation nearly always identified the planted change in the exploratory assessment, and the proposed edit passed the task's tests in 93.1\% of correct rejections (85.1\% on the strongest tests). A developer could thus inspect a specific fault and try a concrete repair. Edits to programs already classed as correct carried a different risk: 17.5\% broke the program or failed to run.

All six models rejected more correct programs under the richer prompt on the audited benchmarks, although the absolute increase was smaller than in the original study. Counterexamples showed why the choice of labels matters: some programs counted as correct had genuine defects. For a team reviewing 100 programs of which 10 are buggy, the estimated cost of asking for an explanation and fix is about 2.0 extra false alarms with adjudicated labels and 4.9 with audited labels. The corresponding adjudicated-label projection is 9.2 candidate repairs that pass the task's tests. A team that uses only verdicts bears the false alarms. A team that reviews edits can weigh that cost against receiving a fault location and a repair to test.

\section*{Statements and Declarations}

\textbf{Funding.} The author received no funding for this work.

\textbf{Competing interests.} The author has no competing interests to declare.

\textbf{Ethics approval and consent.} Not applicable: the study uses public benchmarks and model outputs, and involves no human participants.

\textbf{Data availability.} The replication package contains the audited item set with the reason for every exclusion, every model answer with its request metadata, the oracle runs, the fix and filter outcomes, the analysis scripts that compute every result reported in this article, and the recorded changes to the design after its freeze. It is openly available on Zenodo at \url{https://doi.org/10.5281/zenodo.24634753}. The original study's data and code are available from its authors' repository (commit 642b2bf).

\textbf{Author contributions.} Sabry E. Farrag designed the study, ran the experiments, analysed the data and wrote the article.

\section*{References}

\noindent\hangindent=1.5em\hangafter=1 Austin J, Odena A, Nye M, Bosma M, Michalewski H, Dohan D, Jiang E, Cai C, Terry M, Le Q, Sutton C (2021) Program Synthesis with Large Language Models. arXiv preprint arXiv:2108.07732. \url{https://doi.org/10.48550/arXiv.2108.07732}\par\smallskip
\noindent\hangindent=1.5em\hangafter=1 Bacchelli A, Bird C (2013) Expectations, outcomes, and challenges of modern code review. In: 2013 35th International Conference on Software Engineering (ICSE), pp 712--721. \url{https://doi.org/10.1109/icse.2013.6606617}\par\smallskip
\noindent\hangindent=1.5em\hangafter=1 Cameron AC, Miller DL (2015) A Practitioner's Guide to Cluster-Robust Inference. Journal of Human Resources 50(2):317--372. \url{https://doi.org/10.3368/jhr.50.2.317}\par\smallskip
\noindent\hangindent=1.5em\hangafter=1 Carver JC, Juristo N, Baldassarre MT, Vegas S (2014) Replications of software engineering experiments. Empirical Software Engineering 19(2):267--276. \url{https://doi.org/10.1007/s10664-013-9290-8}\par\smallskip
\noindent\hangindent=1.5em\hangafter=1 Chen B, Zhang F, Nguyen A, Zan D, Lin Z, Lou JG, Chen W (2022) CodeT: Code Generation with Generated Tests. arXiv preprint arXiv:2207.10397. \url{https://doi.org/10.48550/arXiv.2207.10397}\par\smallskip
\noindent\hangindent=1.5em\hangafter=1 Chen M, Tworek J, Jun H, Yuan Q, de Oliveira Pinto HP, Kaplan J, Edwards H, Burda Y, Joseph N, Brockman G, Ray A, Puri R, Krueger G, Petrov M, Khlaaf H, Sastry G, Mishkin P, Chan B, Gray S, Ryder N, Pavlov M, Power A, Kaiser L, Bavarian M, Winter C, Tillet P, Such FP, Cummings D, Plappert M, Chantzis F, Barnes E, Herbert-Voss A, Guss WH, Nichol A, Paino A, Tezak N, Tang J, Babuschkin I, Balaji S, Jain S, Saunders W, Hesse C, Carr AN, Leike J, Achiam J, Misra V, Morikawa E, Radford A, Knight M, Brundage M, Murati M, Mayer K, Welinder P, McGrew B, Amodei D, McCandlish S, Sutskever I, Zaremba W (2021) Evaluating Large Language Models Trained on Code. arXiv preprint arXiv:2107.03374. \url{https://doi.org/10.48550/arXiv.2107.03374}\par\smallskip
\noindent\hangindent=1.5em\hangafter=1 Chen X, Lin M, Sch\"{a}rli N, Zhou D (2023) Teaching Large Language Models to Self-Debug. arXiv preprint arXiv:2304.05128. \url{https://doi.org/10.48550/arXiv.2304.05128}\par\smallskip
\noindent\hangindent=1.5em\hangafter=1 Chen X, Tao Z, Zhang K, Zhou C, Zhang X, Gu W, He Y, Zhang M, Cai X, Zhao H, Jin Z (2025) Revisit Self-Debugging with Self-Generated Tests for Code Generation. In: Proceedings of the 63rd Annual Meeting of the Association for Computational Linguistics (Volume 1: Long Papers), pp 18003--18023. \url{https://doi.org/10.18653/v1/2025.acl-long.881}\par\smallskip
\noindent\hangindent=1.5em\hangafter=1 Cihan U, Haratian V, \.{I}\c{c}\"{o}z A, G\"{u}l MK, Devran \"{O}, Bayendur EF, U\c{c}ar BM, T\"{u}z\"{u}n E (2025a) Automated Code Review in Practice. In: 2025 IEEE/ACM 47th International Conference on Software Engineering: Software Engineering in Practice (ICSE-SEIP), pp 425--436. \url{https://doi.org/10.1109/icse-seip66354.2025.00043}\par\smallskip
\noindent\hangindent=1.5em\hangafter=1 Cihan U, \.{I}\c{c}\"{o}z A, Haratian V, T\"{u}z\"{u}n E (2025b) Evaluating Large Language Models for Code Review. arXiv preprint arXiv:2505.20206. \url{https://doi.org/10.48550/arXiv.2505.20206}\par\smallskip
\noindent\hangindent=1.5em\hangafter=1 Clopper CJ, Pearson ES (1934) The Use of Confidence or Fiducial Limits Illustrated in the Case of the Binomial. Biometrika 26(4):404--413. \url{https://doi.org/10.1093/biomet/26.4.404}\par\smallskip
\noindent\hangindent=1.5em\hangafter=1 Crupi G, Tufano R, Velasco A, Mastropaolo A, Poshyvanyk D, Bavota G (2025) On the Effectiveness of LLM-as-a-Judge for Code Generation and Summarization. IEEE Transactions on Software Engineering 51(8):2329--2345. \url{https://doi.org/10.1109/tse.2025.3586082}\par\smallskip
\noindent\hangindent=1.5em\hangafter=1 Gu J, Jiang X, Shi Z, Tan H, Zhai X, Xu C, Li W, Shen Y, Ma S, Liu H, Wang S, Zhang K, Wang Y, Gao W, Ni L, Guo J (2024) A Survey on LLM-as-a-Judge. arXiv preprint arXiv:2411.15594. \url{https://doi.org/10.48550/arXiv.2411.15594}\par\smallskip
\noindent\hangindent=1.5em\hangafter=1 Guo J, Gu S, Jin M, Lavaei J (2026) Groundability, Not Scale Alone: When Weak Reviewers Can Audit Strong Coding Agents. arXiv preprint arXiv:2610.01023. \url{https://doi.org/10.48550/arXiv.2610.01023}\par\smallskip
\noindent\hangindent=1.5em\hangafter=1 Huang J, Chen X, Mishra S, Zheng HS, Yu AW, Song X, Zhou D (2023) Large Language Models Cannot Self-Correct Reasoning Yet. arXiv preprint arXiv:2310.01798. \url{https://doi.org/10.48550/arXiv.2310.01798}\par\smallskip
\noindent\hangindent=1.5em\hangafter=1 Jiang H, Chen Y, Cao Y, Lee Hy, Tan RT (2025) CodeJudgeBench: Benchmarking LLM-as-a-Judge for Coding Tasks. arXiv preprint arXiv:2507.10535. \url{https://doi.org/10.48550/arXiv.2507.10535}\par\smallskip
\noindent\hangindent=1.5em\hangafter=1 Jin H, Chen H (2025) Uncovering Systematic Failures of LLMs in Verifying Code Against Natural Language Specifications. In: 2025 40th IEEE/ACM International Conference on Automated Software Engineering (ASE), pp 3819--3823. \url{https://doi.org/10.1109/ase63991.2025.00323}\par\smallskip
\noindent\hangindent=1.5em\hangafter=1 Jin H, Chen H (2026) Are LLMs Reliable Code Reviewers? Systematic Overcorrection in Requirement Conformance Judgement. Automated Software Engineering 33(3):90. \url{https://doi.org/10.1007/s10515-026-00638-5}\par\smallskip
\noindent\hangindent=1.5em\hangafter=1 Jin T, Choi Y, Zhu Y, Kang D (2026) Pervasive Annotation Errors Break Text-to-SQL Benchmarks and Leaderboards. arXiv preprint arXiv:2601.08778. \url{https://doi.org/10.48550/arXiv.2601.08778}\par\smallskip
\noindent\hangindent=1.5em\hangafter=1 Kamoi R, Zhang Y, Zhang N, Han J, Zhang R (2024) When Can LLMs Actually Correct Their Own Mistakes? A Critical Survey of Self-Correction of LLMs. Transactions of the Association for Computational Linguistics 12:1417--1440. \url{https://doi.org/10.1162/tacl_a_00713}\par\smallskip
\noindent\hangindent=1.5em\hangafter=1 Kim J (2026) Understanding Regression in Code Self-Refinement: A Study of Refinement Protocols Without Execution Feedback. AI 7(10):416. \url{https://doi.org/10.3390/ai7100416}\par\smallskip
\noindent\hangindent=1.5em\hangafter=1 Lakens D (2017) Equivalence Tests: A Practical Primer for t Tests, Correlations, and Meta-Analyses. Social Psychological and Personality Science 8(4):355--362. \url{https://doi.org/10.1177/1948550617697177}\par\smallskip
\noindent\hangindent=1.5em\hangafter=1 Li Z, Lu S, Guo D, Duan N, Jannu S, Jenks G, Majumder D, Green J, Svyatkovskiy A, Fu S, Sundaresan N (2022) Automating code review activities by large-scale pre-training. In: Proceedings of the 30th ACM Joint European Software Engineering Conference and Symposium on the Foundations of Software Engineering, pp 1035--1047. \url{https://doi.org/10.1145/3540250.3549081}\par\smallskip
\noindent\hangindent=1.5em\hangafter=1 Liang KY, Zeger SL (1986) Longitudinal data analysis using generalized linear models. Biometrika 73(1):13--22. \url{https://doi.org/10.1093/biomet/73.1.13}\par\smallskip
\noindent\hangindent=1.5em\hangafter=1 Lin D, Koppel J, Chen A, Solar-Lezama A (2017) QuixBugs: a multi-lingual program repair benchmark set based on the quixey challenge. In: Proceedings Companion of the 2017 ACM SIGPLAN International Conference on Systems, Programming, Languages, and Applications: Software for Humanity, pp 55--56. \url{https://doi.org/10.1145/3135932.3135941}\par\smallskip
\noindent\hangindent=1.5em\hangafter=1 Lin HY, Liang M, Thongtanunam P, Tantithamthavorn K (2026) Is Agentic Code Review Helpful? Mining Developers' Feedback to CodeRabbit Reviews in the Wild. arXiv preprint arXiv:2607.03316. \url{https://doi.org/10.48550/arXiv.2607.03316}\par\smallskip
\noindent\hangindent=1.5em\hangafter=1 Liu J, Xia CS, Wang Y, Zhang L (2023) Is Your Code Generated by ChatGPT Really Correct? Rigorous Evaluation of Large Language Models for Code Generation. arXiv preprint arXiv:2305.01210. \url{https://doi.org/10.48550/arXiv.2305.01210}\par\smallskip
\noindent\hangindent=1.5em\hangafter=1 Madaan A, Tandon N, Gupta P, Hallinan S, Gao L, Wiegreffe S, Alon U, Dziri N, Prabhumoye S, Yang Y, Gupta S, Majumder BP, Hermann K, Welleck S, Yazdanbakhsh A, Clark P (2023) Self-Refine: Iterative Refinement with Self-Feedback. arXiv preprint arXiv:2303.17651. \url{https://doi.org/10.48550/arXiv.2303.17651}\par\smallskip
\noindent\hangindent=1.5em\hangafter=1 Mahmud Y (2026) RTLReviewBench: LLM Hardware Reviewers Rarely Miss Bugs but Routinely Reject Correct RTL. arXiv preprint arXiv:2611.21678. \url{https://doi.org/10.48550/arXiv.2611.21678}\par\smallskip
\noindent\hangindent=1.5em\hangafter=1 Mao X, Qian R, Wang L, Yi X, Li M, Chen L, Gao Y, Zheng X, Wang C (2026) "You're Right, Let Me Fix It": How LLM Agents Damage Correct Work When Falsely Accused. arXiv preprint arXiv:2609.32616. \url{https://doi.org/10.48550/arXiv.2609.32616}\par\smallskip
\noindent\hangindent=1.5em\hangafter=1 McAleese N, Pokorny RM, Uribe JFC, Nitishinskaya E, Trebacz M, Leike J (2024) LLM Critics Help Catch LLM Bugs. arXiv preprint arXiv:2407.00215. \url{https://doi.org/10.48550/arXiv.2407.00215}\par\smallskip
\noindent\hangindent=1.5em\hangafter=1 Moon J, Hwang Y, Lee D, Kang T, Kim Y, Jung K (2025) Don't Judge Code by Its Cover: Exploring Biases in LLM Judges for Code Evaluation. arXiv preprint arXiv:2505.16222. \url{https://doi.org/10.48550/arXiv.2505.16222}\par\smallskip
\noindent\hangindent=1.5em\hangafter=1 Muennighoff N, Liu Q, Zebaze A, Zheng Q, Hui B, Zhuo TY, Singh S, Tang X, von Werra L, Longpre S (2023) OctoPack: Instruction Tuning Code Large Language Models. arXiv preprint arXiv:2308.07124. \url{https://doi.org/10.48550/arXiv.2308.07124}\par\smallskip
\noindent\hangindent=1.5em\hangafter=1 Northcutt CG, Athalye A, Mueller J (2021) Pervasive Label Errors in Test Sets Destabilize Machine Learning Benchmarks. arXiv preprint arXiv:2103.14749. \url{https://doi.org/10.48550/arXiv.2103.14749}\par\smallskip
\noindent\hangindent=1.5em\hangafter=1 Olausson TX, Inala JP, Wang C, Gao J, Solar-Lezama A (2023) Is Self-Repair a Silver Bullet for Code Generation? arXiv preprint arXiv:2306.09896. \url{https://doi.org/10.48550/arXiv.2306.09896}\par\smallskip
\noindent\hangindent=1.5em\hangafter=1 Ralph P, bin Ali N, Baltes S, Bianculli D, Diaz J, Dittrich Y, Ernst N, Felderer M, Feldt R, Filieri A, de Fran\c{c}a BBN, Furia CA, Gay G, Gold N, Graziotin D, He P, Hoda R, Juristo N, Kitchenham B, Lenarduzzi V, Mart\'{\i}nez J, Melegati J, Mendez D, Menzies T, Molleri J, Pfahl D, Robbes R, Russo D, Saarim\"{a}ki N, Sarro F, Taibi D, Siegmund J, Spinellis D, Staron M, Stol K, Storey MA, Taibi D, Tamburri D, Torchiano M, Treude C, Turhan B, Wang X, Vegas S (2020) Empirical Standards for Software Engineering Research. arXiv preprint arXiv:2010.03525. \url{https://doi.org/10.48550/arXiv.2010.03525}\par\smallskip
\noindent\hangindent=1.5em\hangafter=1 Sharma M, Tong M, Korbak T, Duvenaud D, Askell A, Bowman SR, Cheng N, Durmus E, Hatfield-Dodds Z, Johnston SR, Kravec S, Maxwell T, McCandlish S, Ndousse K, Rausch O, Schiefer N, Yan D, Zhang M, Perez E (2023) Towards Understanding Sycophancy in Language Models. arXiv preprint arXiv:2310.13548. \url{https://doi.org/10.48550/arXiv.2310.13548}\par\smallskip
\noindent\hangindent=1.5em\hangafter=1 Song X, Haque H, Somasundaram J, Turner R, Zhu B, Zhao L (2026) CIVER: Causal Intervention Verification for Reliable Code Review. arXiv preprint arXiv:2610.36010. \url{https://doi.org/10.48550/arXiv.2610.36010}\par\smallskip
\noindent\hangindent=1.5em\hangafter=1 Tong W, Zhang T (2024) CodeJudge: Evaluating Code Generation with Large Language Models. In: Proceedings of the 2024 Conference on Empirical Methods in Natural Language Processing, pp 20032--20051. \url{https://doi.org/10.18653/v1/2024.emnlp-main.1118}\par\smallskip
\noindent\hangindent=1.5em\hangafter=1 Tufano R, Masiero S, Mastropaolo A, Pascarella L, Poshyvanyk D, Bavota G (2022) Using pre-trained models to boost code review automation. In: Proceedings of the 44th International Conference on Software Engineering, pp 2291--2302. \url{https://doi.org/10.1145/3510003.3510621}\par\smallskip
\noindent\hangindent=1.5em\hangafter=1 Wang D, Zhao J, Pei H, Tan S, Zha S (2024) Fine-tuning Language Models for Joint Rewriting and Completion of Code with Potential Bugs. In: Findings of the Association for Computational Linguistics ACL 2024, pp 15854--15868. \url{https://doi.org/10.18653/v1/2024.findings-acl.938}\par\smallskip
\noindent\hangindent=1.5em\hangafter=1 Wang R, Wang X, Gao C, Chong CY, Xia X, Liao Q (2025) Benchmarking Code LLM-as-a-Judge via Constrained Perturbation and Calibration. arXiv preprint arXiv:2512.20159. \url{https://doi.org/10.48550/arXiv.2512.20159}\par\smallskip
\noindent\hangindent=1.5em\hangafter=1 Wang Y, Pradel M, Liu Z (2026a) Are "Solved Issues" in SWE-bench Really Solved Correctly? An Empirical Study. In: Proceedings of the 2026 IEEE/ACM 48th International Conference on Software Engineering, pp 169--181. \url{https://doi.org/10.1145/3744916.3764576}\par\smallskip
\noindent\hangindent=1.5em\hangafter=1 Xiang Z, Zhang Y, Zhang Y, Xu H (2026) Cross-Model LLM Code Review: Should you use Claude to review Codex or vice versa? arXiv preprint arXiv:2607.21656. \url{https://doi.org/10.48550/arXiv.2607.21656}\par\smallskip
\noindent\hangindent=1.5em\hangafter=1 Yadavally A, Nguyen H, Callot L, Guinet G (2025) Large Language Model Critics for Execution-Free Evaluation of Code Changes. arXiv preprint arXiv:2501.16655. \url{https://doi.org/10.48550/arXiv.2501.16655}\par\smallskip
\noindent\hangindent=1.5em\hangafter=1 Ye H, Martinez M, Durieux T, Monperrus M (2021) A comprehensive study of automatic program repair on the QuixBugs benchmark. Journal of Systems and Software 171:110825. \url{https://doi.org/10.1016/j.jss.2020.110825}\par\smallskip
\noindent\hangindent=1.5em\hangafter=1 Zhan S, Wang X, Wei D, Xiang X (2026) How Reliable Is LLM-as-Judge for Patch Correctness Assessment? An Empirical Study. In: LIPIcs, Volume 394, ESEM 2026, pp 21:1--21:20. \url{https://doi.org/10.4230/LIPICS.ESEM.2026.21}\par\smallskip
\noindent\hangindent=1.5em\hangafter=1 Zheng L, Chiang WL, Sheng Y, Zhuang S, Wu Z, Zhuang Y, Lin Z, Li Z, Li D, Xing EP, Zhang H, Gonzalez JE, Stoica I (2023) Judging LLM-as-a-Judge with MT-Bench and Chatbot Arena. arXiv preprint arXiv:2306.05685. \url{https://doi.org/10.48550/arXiv.2306.05685}\par\smallskip

\end{document}